\documentclass[reprint, superscriptaddress, amsmath, amssymb, aps, prl]{revtex4-2}

\usepackage{graphicx}
\usepackage{dcolumn}
\usepackage{bm}
\usepackage{comment}
\usepackage{amsmath, amssymb}
\usepackage{bbold}
\usepackage{xcolor}
\usepackage{hyperref}

\begin{document}

\preprint{APS/123-QED}

\title{Orbital-selective spin-triplet superconductivity in infinite-layer LaNiO$_{2}$} 

\author{Fabian Jakubczyk}
\email{fabian.jakubczyk@tu-dresden.de}
\affiliation{Institute of Theoretical Physics, Technische Universität Dresden, 01069 Dresden, Germany}
\affiliation{Würzburg-Dresden Cluster of Excellence ct.qmat, Germany}

\author{Armando Consiglio}
\affiliation{Würzburg-Dresden Cluster of Excellence ct.qmat, Germany}
\affiliation{Institut für Theoretische Physik und Astrophysik, Universität Würzburg, 97074 Würzburg, Germany}
\affiliation{Istituto Officina dei Materiali, Consiglio Nazionale delle Ricerche, Trieste I-34149, Italy}

\author{Domenico Di Sante}
\affiliation{Department of Physics and Astronomy,
University of Bologna, 40127 Bologna, Italy}

\author{Ronny Thomale}
\affiliation{Würzburg-Dresden Cluster of Excellence ct.qmat, Germany}
\affiliation{Institut für Theoretische Physik und Astrophysik, Universität Würzburg, 97074 Würzburg, Germany}

\author{Carsten Timm}
\affiliation{Institute of Theoretical Physics, Technische Universität Dresden, 01069 Dresden, Germany}
\affiliation{Würzburg-Dresden Cluster of Excellence ct.qmat, Germany}

\date{\today}

\begin{abstract}
The discovery of superconductivity in infinite-layer nickelates has ignited stark interest within the scientific community, particularly regarding its likely unconventional origin. 
Conflicting magnetotransport measurements report either isotropic or anisotropic suppression of superconductivity in an external magnetic field, with distinct implications for the nature of superconducting order. 
In order to ensure a most suited model subject to subsequent many-body analysis, we develop a first-principles-guided minimal theory including Ni $d_{x^2-y^2}$, La $d_{3z^2-r^2}$, and La $d_{xy}$ orbitals. 
Amended by the consideration of orbital-selective pairing formation, which emphasises the correlation state of the Ni $3d_{x^2-y^2}$ orbital, we calculate the superconducting ordering susceptibility mediated by spin fluctuations. 
We find a parametric competition between even-parity \textit{d}-wave and, in contrast to previous studies, odd-parity \textit{p}-wave pairing, which becomes favorable through a large quasiparticle weight renormalization for Ni $3d_{x^2-y^2}$ electrons. 
Our findings not only shed light on the distinctiveness of LaNiO$_{2}$ as compared to cuprate superconductors or nickelates of different rare-earth composition but also suggest similarities to other candidate odd-parity superconductors.
\end{abstract}

\maketitle

\textit{Introduction.}---Since the initial discovery of superconductivity in Sr-doped NdNiO$_{2}$ thin films \cite{li_superconductivity_2019}, the family of nickelate superconductors has kept on growing and by now also includes infinite-layer (IL) La$_{1-x}$Sr$_{x}$NiO$_{2}$, Pr$_{1-x}$Sr$_{x}$NiO$_{2}$, and La$_{1-x}$Ca$_{x}$NiO$_{2}$ \cite{osada_nickelate_2021, zeng_superconductivity_2022}. 
Most intriguingly, some groups report the onset of superconductivity even in undoped IL nickelates, suggesting that the eventual ground state of sufficiently clean parent nickelates is a superconducting (SC) state \cite{osada_nickelate_2021,fowlie_intrinsic_2022, sun_evidence_2023, parzyck_superconductivity_2024, sahib_superconductivity_2024}.
Despite being isostructural to high $T_{c}$ cuprates, the extend of similarity between these materials remains partially unclear, in spite of numerous studies addressing this issue \cite{lee_infinite-layer_2004, botana_similarities_2020, lechermann_late_2020, anisimov_electronic_1999, hepting_electronic_2020}. 

A fundamental subject of discussion is the symmetry of the SC order parameter (OP).
On the one hand, magnetotransport measurements indicate isotropic Pauli-limited behavior with singlet pairing and even parity in the IL nickelate Nd$_{0.775}$Sr$_{0.225}$NiO$_{2}$ \cite{wang_isotropic_2021}. This scenario was supported by many theoretical investigations, mostly pointing at \textit{d}-wave order like in the cuprates \cite{sakakibara_model_2020, wu_robust_2020, kitatani_nickelate_2020, adhikary_orbital-selective_2020, xie_microscopic_2022, lu_two-orbital_2022}. On the other hand, evidence for anisotropic superconductivity that violates Pauli limiting and potential spin-triplet pairing was reported in La-based nickelate thin films \cite{sun_evidence_2023, chow_pauli-limit_nodate, wei_large_2023}. Likewise, this anisotropic limiting behavior for different magnetic-field orientations has recently been observed for free-standing IL nickelate membranes \cite{yan_superconductivity_2024}. Variations in the upper critical field within the nickelate family were ascribed to the different rare-earth (RE) elements, i.e., the magnetic character of the 4\textit{f} electrons for the Nd$^{3+}$ Kramers doublet as opposed to their absence in La$^{3+}$, or the nonmagnetic singlet ground state of Pr$^{3+}$ \cite{wang_effects_2023}. A rare-earth-specific order parameter is also observed by London-penetration-depth measurements \cite{harvey_evidence_2022, chow_pairing_2023}.

In this Letter, we address the open question of gap structure and potential odd-parity pairing in IL LaNiO$_{2}$. 
The largest deviation from the cuprates as well as the strongest effective-mass enhancement can be expected in the absence of doping.
A direct comparison between for instance NdNiO$_{2}$ and LaNiO$_{2}$ reveals that this effect is particularly strong in the lanthanum compound \cite{kitatani_nickelate_2020}. 
Moreover, LaNiO$_{2}$ does not host any 4\textit{f} electrons, who may influence the SC pairing \cite{choi_role_2020}. Hybridization with 4\textit{f} electrons is expected to be non-negligible in Nd and Pr nickelates \cite{zhang_magnetic_2021, choi_role_2020, jiang_electronic_2019, bandyopadhyay_superconductivity_2020}. To resolve this puzzle, we adopt the perspective of spin-fluctuation-mediated unconventional superconductivity \cite{scalapino_common_2012} and incorporate the effect of increased correlations via the orbital-selective approach known from iron-based superconductors \cite{yamakawa_nematicity_2016, sprau_discovery_2017, kreisel_orbital_2017}. Our calculations reveal that spin-triplet pairing can indeed be realized in LaNiO$_{2}$, once an enhanced effective mass of the Ni 3$d_{x^2-y^2}$ electrons is taken into account.

\textit{Low-energy electronic structure.}---In contrast to the parent compounds of the cuprates, which exhibit long-range antiferromagnetic order, clear signatures of long-range magnetism are lacking for IL nickelates, suggesting a paramagnetic ground state, in particular for LaNiO$_{2}$ \cite{lin_universal_2022, ortiz_magnetic_2022, zhao_intrinsic_2021}. 
Magnetism is often triggered by strong electronic correlations, which indeed exist in IL nickelates \cite{botana_similarities_2020, kitatani_nickelate_2020, lechermann_assessing_2022, kang_optical_2021}.  
In fact, stable magnetic order was proposed based on \textit{ab-initio} calculations \cite{zhang_magnetic_2021, kapeghian_electronic_2020, wang_hunds_2020, botana_similarities_2020}. 
The unexpected absence of long-range magnetic order might be explained by the coupling to itinerant RE electrons, i.e., by the destruction of long-range order via self-doped holes, an effect not well captured by mean-field calculations \cite{botana_similarities_2020, lu_magnetic_2021}. 
Nevertheless, nonlocal spin fluctuations have been observed in IL nickelates \cite{ortiz_magnetic_2022, wu_robust_2020, sakakibara_model_2020, kitatani_nickelate_2020, lu_magnetic_2021, zhao_intrinsic_2021, kreisel_superconducting_2022, worm_spin_2023, zhang_magnetic_2021}, which makes spin-fluctuation-mediated superconductivity the most reasonable mechanism. 
Furthermore, muon spin rotation ($\mu$SR) has recently demonstrated intrinsic short-range magnetic order, also in the superconducting state \cite{fowlie_intrinsic_2022}. 
Scanning-SQUID measurements found paramagnetic response in La$_{0.85}$Sr$_{0.15}$NiO$_{2}$, together with an inhomogeneous ferromagnetic background independent of the RE element \cite{shiscanning2024}. 
However, short-range magnetic order is not expected to significantly alter the electronic band structure since magnetic effects lacking long-range periodicity are averaged out. This results in a Fermi surface that is very similar to the one in the absence of magnetism. The excellent match between recent ARPES measurements on Sr-doped LaNiO$_{2}$ \cite{sun_electronic_2024} and band-structure calculations of non-magnetic LaNiO$_{2}$ \cite{sakakibara_model_2020, wang_hunds_2020, kang_optical_2021, hepting_electronic_2020, sun_electronic_2024} supports this conclusion.

In non-magnetic LaNiO$_{2}$, the principal contribution at the Fermi level arises from a band primarily governed by Ni 3$d_{x^2-y^2}$ electrons. This results in a large hole-like and mostly two-dimensional Fermi surface (FS) similar to the cuprates.
In addition, another hybridized band of mixed Ni 3\textit{d} and La 5\textit{d} character creates small ellipsoidal, electron-like pockets around the $\Gamma$ and A high-symmetry points of the Brillouin zone, indicating the multiorbital and three-dimensional (3D) character of LaNiO$_{2}$ \cite{wang_hunds_2020, werner_nickelate_2020, hepting_electronic_2020}. These two electron pockets enable self-hole-doping of the large $d_{x^2-y^2}$ Fermi sheet, a process with major involvement of La 5\textit{d} states \cite{lechermann_assessing_2022, botana_similarities_2020}. Especially La 5$d_{z^2}$ and La 5$d_{xy}$ orbitals are known to have significant weight on the FS \cite{botana_similarities_2020, hepting_electronic_2020, sakakibara_model_2020}, where $d_{z^2}$ is a shorthand notation for $d_{3z^2-r^2}$. Furthermore, these partially occupied electron pockets might give rise to the negative Hall coefficient, to the slightly metallic behavior of the resistivity at high temperatures, and possibly even to the superconductivity in undoped IL nickelates \cite{osada_nickelate_2021, zeng_superconductivity_2022, sakakibara_model_2020}.

Hole-doping the system for instance with Sr induces changes in the electronic structure. It reduces the self-doping effect by first shifting the $\Gamma$ pocket above $E_{F}$, followed by the pocket around A in the overdoped region \cite{kitatani_nickelate_2020, botana_similarities_2020, krishna_effects_2020, leonov_lifshitz_2020, sun_electronic_2024}. Furthermore, doping reduces the charge-transfer energy \cite{krishna_effects_2020}, i.e., overall increases the cuprate-like character of the nickelates.  
However, the impact on LaNiO$_{2}$ is expected to be less pronounced compared to Nd-/Pr-based compounds regarding charge carriers, which is evidenced by the persistent negative Hall coefficient of La$_{1-x}$Sr$_{x}$NiO$_{2}$ even upon doping \cite{osada_nickelate_2021, zeng_superconductivity_2022}. 

\begin{figure}[t!]
\centering
\includegraphics[width=.48\textwidth]{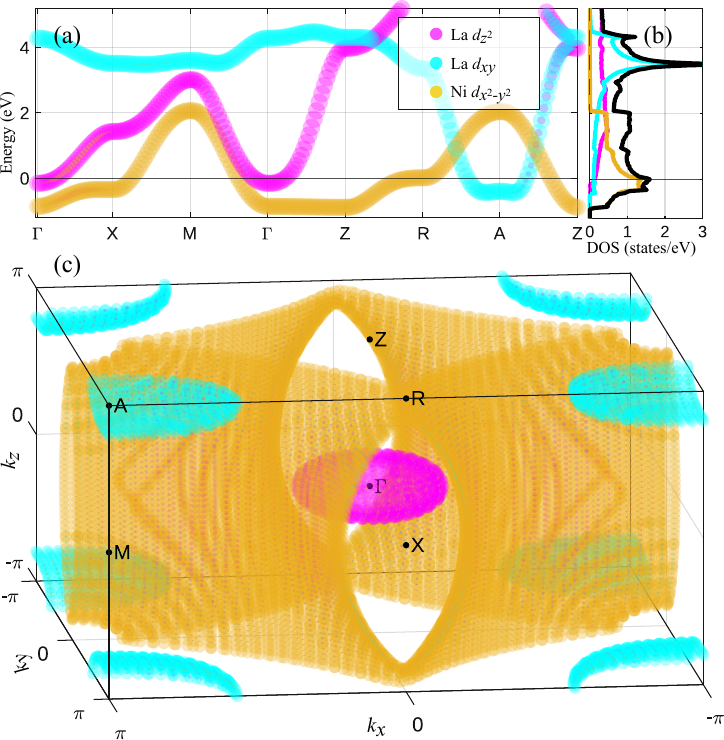}
\caption{(a) Band structure, (b) density of states, and (c) 3D Fermi surface including the orbital contributions relevant for the low-energy description of prisitine LaNiO$_{2}$ without inclusion of renormalized quasiparticle weights.
Magenta, cyan, and yellow refer to the La $d_{z^2}$, La $d_{xy}$, and Ni $d_{x^2-y^2}$ orbitals, respectively.}
\label{fig:bands_DOS_3D_FS}
\end{figure}

In our effective low-energy Hamiltonian, based on the non-magnetic electronic structure of the system, we thus include Ni $d_{x^2-y^2}$, La $d_{z^2}$, and La $d_{xy}$ orbitals and introduce the operator $\psi_{\sigma}^{\dagger} (\mathbf{k})= [ c_{1\sigma}^{\dagger} (\mathbf{k}), c_{2\sigma}^{\dagger}(\mathbf{k}), c_{3\sigma}^{\dagger} (\mathbf{k}) ]$ to describe the given multiband system. Here, $c_{l \sigma}^{\dagger}(\mathbf{k})$ is the fermionic creation operator, where $\sigma$ and $l$ denote the spin and orbital index, with $l=1,2,3$ referring to the La 5$d_{z^2}$, La 5$d_{xy}$, and Ni 3$d_{x^2-y^2}$ orbital, respectively. We thus write the non-interacting Hamiltonian as
\begin{align}
    H_{0}
    = \sum_{\mathbf{k} \sigma} \psi_{\sigma}^{\dagger}(\mathbf{k}) h (\mathbf{k}) \psi_{\sigma}(\mathbf{k}),
\label{eq:Hamiltonian_TB_momentum_general}
\end{align}
where $h (\mathbf{k})$ is constructed utilizing hopping amplitudes obtained from the Wannier downfolding of the band structure derived from density functional theory (DFT), wherein spin polarization is not taken into account.
Thereby, we capture the important characteristics, as well as the distribution of orbital characters, of the non-magnetic FS in agreement with previous DFT \cite{zhang_magnetic_2021, sakakibara_model_2020, lane_competing_2023} and alternative approaches such as DFT+$U$ \cite{botana_similarities_2020, liu_electronic_2020}, DFT + dynamical mean-field theory (DMFT) \cite{wang_hunds_2020, kang_optical_2021} or the \textit{GW} approximation \cite{olevano_ab_2020}. 
Details on the \textit{ab-initio} calculations including further justification of the non-magnetic electronic structure, as well as the Hamiltonian $h (\mathbf{k})$ are given in the Supplemental Material (SM) \cite{sup_mat}.
In Fig.\ \ref{fig:bands_DOS_3D_FS}, we plot the orbitally resolved single-particle band structure, as well as the corresponding 3D FS and density of states (DOS) of our minimal model. Here, La 5$d_{z^2}$ weight is located mostly around $\Gamma$, while the La 5$d_{xy}$ electrons have a major contribution around the A point.

\textit{Strong correlations and magnetic response.}---Orbitally differentiated correlations play an important role in IL nickelates. In particular the Ni 3$d_{x^2-y^2}$ electrons are highly correlated, suggesting that they are in the vicinity of a Mott critical regime \cite{wang_hunds_2020}. This is evidenced by an enhanced effective mass and thus reduced quasiparticle weight $Z$, which is significantly smaller for the Ni $d_{x^2-y^2}$ orbital compared to the other 3\textit{d} orbitals \cite{xie_microscopic_2022, kang_optical_2021, kitatani_nickelate_2020}. 
In our work, we therefore focus on the effect of a highly correlated Ni $d_{x^2-y^2}$ orbital. Previous studies have reported specific values of either the effective mass or the renormalization factor $Z$ ($m^*/m = 1/Z$) in LaNiO$_{2}$, ranging from $m^*/m \approx 2.81$ ($Z_{d_{x^2-y^2}} \approx 0.35$) \cite{kang_optical_2021, wang_hunds_2020}, over $m^*/m \approx 4.1$ ($Z_{d_{x^2-y^2}} \approx 0.24$) \cite{ryee_induced_2020} to $m^*/m \approx 5.5$ ($Z_{d_{x^2-y^2}} \approx 0.18$) \cite{kitatani_nickelate_2020} 
and $m^*/m \approx 7$ ($Z_{d_{x^2-y^2}} \approx 0.14$) \cite{si_topotactic_2020} or even lower ($Z_{d_{x^2-y^2}} \approx 0$) \cite{xie_microscopic_2022}. 
Doping would shift the Ni $d_{x^2-y^2}$ orbital away from half filling, where electronic correlations are strongest, leading to a reduced mass enhancement \cite{kitatani_nickelate_2020, xie_microscopic_2022, ryee_induced_2020, wang_hunds_2020}.

We take this correlation-induced renormalization into account by means of the orbital selective ansatz \cite{yamakawa_nematicity_2016, kreisel_orbital_2017}. In this ansatz, quasiparticles in orbital $l$ are weighted by a factor $\sqrt{Z_{l}}$, i.e., $c_{l}^{\dagger}(\mathbf{k}) \rightarrow \sqrt{Z_{l}} c_{l}^{\dagger}(\mathbf{k})$ and the Green's function in the orbital basis becomes
\begin{align}
    \tilde{G}_{ll'}(\mathbf{k}, \omega_{n}) = 
    \sqrt{Z_{l}Z_{l'}}\sum_{\mu} \frac{a_{\mu}^{l}(\mathbf{k}) a_{\mu}^{l'*}(\mathbf{k}) }{i\omega_{n} - E_{\mu}(\mathbf{k})},
\label{eq:Greens_function_correlated}
\end{align}
where $E_{\mu}(\mathbf{k})$ is the eigenenergy of band $\mu$. 
Subsequently, the bare susceptibility in orbital space $\chi^{0}_{l_{1}l_{2}l_{3}l_{4}}$ requires a straightforward multiplication by the quasiparticle weights to derive the corresponding quantity 
\begin{align}
    \tilde{\chi}^{0}_{l_{1}l_{2}l_{3}l_{4}} (\mathbf{q}, \omega) = \sqrt{Z_{l_{1}}Z_{l_{2}}Z_{l_{3}}Z_{l_{4}}} 
    ~ \chi^{0}_{l_{1}l_{2}l_{3}l_{4}} (\mathbf{q}, \omega)  
\end{align}
in the correlated system \cite{yamakawa_nematicity_2016, kreisel_orbital_2017, kreisel_superconducting_2022}.   

In the next step, we introduce local electron-electron interactions via a multiorbital Hubbard-Hund Hamiltonian
\begin{align}
    H_{\text{int}} = \;\;\; &{U}_{\text{Ni}} \sum_{i} n_{i\uparrow} n_{i\downarrow} \nonumber \\
    + \; &{U}_{\text{La}} \sum_{i} n_{il\uparrow} n_{il\downarrow}
    + {U}_{\text{La}}' \sum_{i, l'<l} n_{il} n_{il'} \nonumber \\ 
    + \; &{J}_{\text{La}} \sum_{i, l'<l} \sum_{\sigma \sigma'} c_{il\sigma}^{\dagger} c_{il'\sigma'}^{\dagger} c_{il\sigma'} c_{il'\sigma} \nonumber \\
    + \; &{J}_{\text{La}}' \sum_{i, l'\neq l} c_{il\uparrow}^{\dagger} c_{il\downarrow}^{\dagger} c_{il'\downarrow} c_{il'\uparrow}, 
    \label{eq:Hamiltonian_Hubbard_Hund}
\end{align}
where the coupling constants ${U}$, ${U}'$, $J$, and ${J}'$ denote the intraorbital Coulomb, interorbital Coulomb, Hund's, and pair-hopping interactions, respectively.
The orbital indices $l,l' \in (1,2) $ run over the La 5\textit{d} orbitals $d_{z^2}$ and $d_{xy}$. 
Throughout our calculations, we apply the relations $U'=U-(J+J')$ and $J'=J$ between the coupling constants, which are satisfied in various orbital degenerate models and if orbital wave functions can be chosen real \cite{kubo_pairing_2007, tang_theory_1998}. 
Guided by prior research \cite{sakakibara_model_2020, kang_infinite-layer_2023, wang_hunds_2020, kang_optical_2021, xie_microscopic_2022, ryee_induced_2020}, we estimate the interaction strengths in this material by $U_{\text{La}}/ U_{\text{Ni}} = 0.5$ and consider the values $J_{\text{La}}/U_{\text{La}} = 0.2$, as well as $J_{\text{La}}/U_{\text{La}} = 0.1$.
The inclusion of on-site Hund’s coupling $J$, i.e., of intra-atomic exchange interactions is expected to be an important ingredient for capturing the correlation physics. LaNiO$_{2}$ shows characteristic Hund’s metal signatures such as the unexpected absence of long-range magnetism, the importance of high-spin configurations, metalliticity, orbital differentiation, and a decrease of $Z_{d_{x^2-y^2}}$ with stronger Hund's coupling $J$ \cite{wang_hunds_2020, kang_optical_2021, kang_infinite-layer_2023, xie_microscopic_2022}. 
These interactions are included in the calculation of susceptibilities within the random phase approximation (RPA). 
Note that in RPA the spin susceptibility systematically diverges once the interaction exceeds a critical value $U_{c}$ and indicates a spin-density-wave instability, below which superconductivity emerges induced by spin fluctuations. 

\begin{figure}[t]
\centering
\includegraphics[width=.48\textwidth]{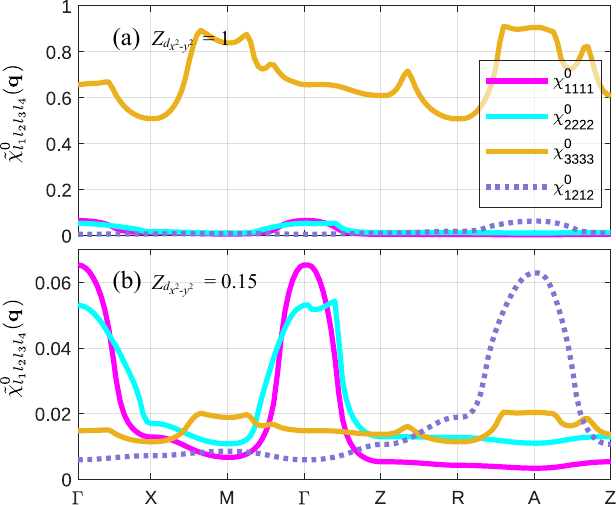}
\caption{Static bare susceptibilities $\tilde{\chi}^{0}_{l_{1}l_{2}l_{3}l_{4}} (\mathbf{q}, \omega = 0) $ of parent LaNiO$_{2}$ for (a) $Z_{d_{x^2-y^2}} = 1$ and (b) $Z_{d_{x^2-y^2}} = 0.15$. Magenta, cyan, and yellow refer to the pure La $d_{z^2}$, La $d_{xy}$ and Ni $d_{x^2-y^2}$ orbital contributions, respectively. The interorbital bare susceptibility $\tilde{\chi}^{0}_{1212}$ is shown in purple. We use $80 \times 80 \times 80$ \textbf{k}-points for momentum space integration.}
\label{fig:sus_path_orbitally}
\end{figure}

\begin{figure*}[t]
\centering
\includegraphics[width=.98\textwidth]{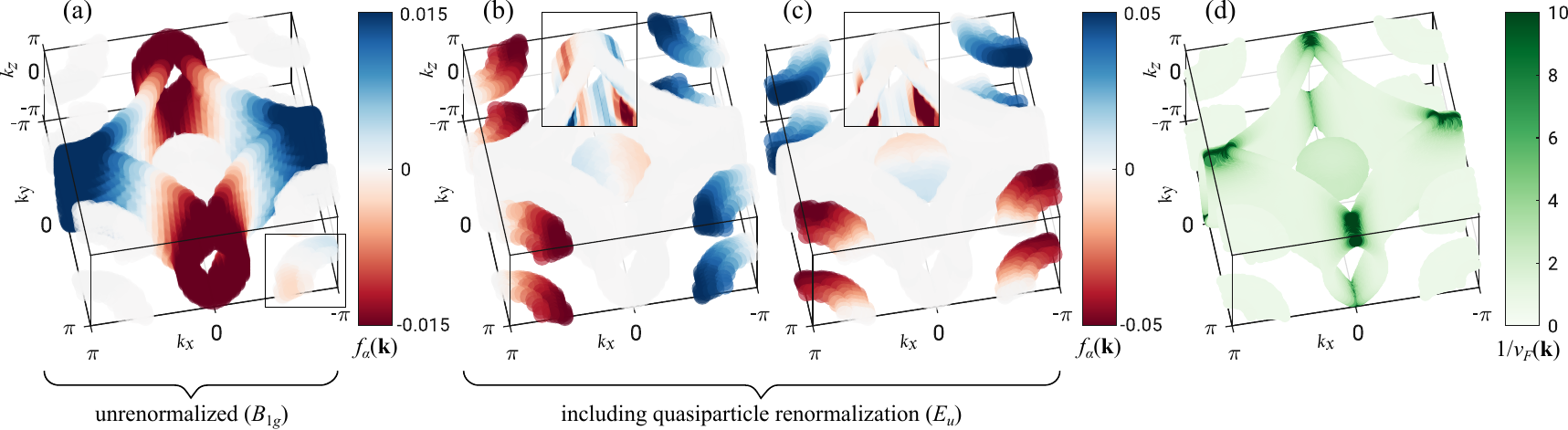}
\caption{Leading pairing structure of LaNiO$_{2}$ (a) without and (b), (c) with inclusion of quasiparticle renormalization.
The gap symmetry changes from (a) an even-parity $d$-wave to (b), (c) a degenerate odd-parity $p$-wave solution. 
A color scale highlighting smaller amplitudes with maximum value of $\pm 1 \times 10^{-4}$ was chosen for the insets to reveal, e.g., in panels (b), (c), the horizontal line nodes on the large FS.
(d) Inverse Fermi-velocity $1/v_{F}(\mathbf{k}) = 1/|\nabla_{\mathbf{k}} E_{\nu} (\mathbf{k})|$ giving a measure of the DOS. 
A total of 8504 points is taken for the 3D gap-structure calculations and we use $60 \times 60 \times 60$ \textbf{k}-points for momentum space integration.}
\label{fig:gap_leading_and_inverse_Fermi-velocity}
\end{figure*}  

In Figs.$~$\ref{fig:sus_path_orbitally}(a) and \ref{fig:sus_path_orbitally}(b), the non-interacting susceptibilities $\tilde{\chi}^{0}_{l_{1}l_{2}l_{3}l_{4}} (\mathbf{q}, 0) $ are shown separately for the orbital contributions $l=1,2,3$ referring to La 5$d_{z^2}$, La 5$d_{xy}$, and Ni 3$d_{x^2-y^2}$, respectively. Figure$~$\ref{fig:sus_path_orbitally}(a) addresses the unrenormalized case, with the main peaks located around the M and A points, similar to the cuprates \cite{romer_pairing_2015}. These magnetic fluctuations are dominated by Ni $d_{x^2-y^2}$ electrons and strongly enhanced upon introducing interactions at the RPA level. In contrast, the diagonal contributions of the La $d_{z^2}$ and $d_{xy}$ orbital reach their maximum around $\Gamma$. Moreover, we observe a sizable interorbital $\tilde{\chi}^{0}_{1212}$ feature around the A point. 

Physically reasonable reduction of the Ni $d_{x^2-y2}$ quasiparticle weight results in a clear dominance of fluctuations stemming from the RE orbitals, as can be seen in Fig.\ \ref{fig:sus_path_orbitally}(b), where $Z_{d_{x^2-y^2}} = 0.15$. 
Interactions predominantly lead to an enhancement of the diagonal La 5\textit{d} peaks around $\Gamma$ but also of interorbital $\tilde{\chi}_{1122}^{\text{RPA}, \,s}$ and $\tilde{\chi}_{1221}^{\text{RPA}, \,s}$ contributions around $\Gamma$ and A, respectively. The latter are not notable on the bare level but in particular the $\tilde{\chi}_{1122}^{\text{RPA}, \,s}$ component diverges together with the intraorbital spin susceptibilities $\tilde{\chi}_{1111}^{\text{RPA}, \,s}$ and $\tilde{\chi}_{2222}^{\text{RPA}, \,s}$ upon approaching $U_{c}$. 
Previous calculations of the LaNiO$_{2}$ static weak-coupling spin susceptibility for $\mathbf{q} = (q_{x}, q_{y}, 0)$ with converged DFT + DMFT Green's functions similarly resulted in a maximum near $\Gamma$ \cite{lechermann_assessing_2022}. 
Such a transition from primarily antiferromagnetic to ferromagnetic fluctuations stresses the necessity of reevaluating the SC pairing symmetries in the regime of strongly renormalized Ni 3\textit{d} quasiparticle weights \cite{kitamura_spin-triplet_2024} .

\textit{Spin-fluctuation pairing.}---Magnetic fluctuations are seen as a strong candidate for mediating superconductivity in IL nickelates \cite{ortiz_magnetic_2022, wu_robust_2020, sakakibara_model_2020, kitatani_nickelate_2020, lu_magnetic_2021, zhao_intrinsic_2021, kreisel_superconducting_2022, worm_spin_2023, zhang_magnetic_2021}. In order to compute the SC pairing interactions, we thus employ a formalism based on spin-fluctuations \cite{berk_effect_1966, graser_near-degeneracy_2009, kreisel_spin_2013} with the adjustment of additional quasiparticle renormalization when projecting the pairing interaction from orbital to band space \cite{kreisel_orbital_2017, kreisel_superconducting_2022}. 
For this purpose, we solve the linearized gap equation 
\begin{align}
    \lambda_{\alpha} f_{\alpha}(\mathbf{k}) = 
    \sum_{j} \oint_{C_{j}} \frac{\text{d}^2 k_{\|}'}{(2\pi)^2 v_{F}(\mathbf{k}')} \tilde{\Gamma}_{ij}(\mathbf{k},\mathbf{k'}) f_{\alpha}(\mathbf{k}'),
\label{eq:linearized_gap_equation}
\end{align}
the eigenvalues of which determine the pairing strength $\lambda_{\alpha}$ for the various pairing channels $\alpha$ \cite{scalapino_d_1986}. The largest eigenvalue results in the highest transition temperature \cite{graser_near-degeneracy_2009}, while its eigenfunction $f_{\alpha}(\mathbf{k})$ identifies the gap symmetry. 
$\tilde{\Gamma}_{ij}(\mathbf{k},\mathbf{k'})$ represents the renormalized form of the effective multiorbital pair scattering vertex and momenta $\mathbf{k} \in C_{i}$, $\mathbf{k}' \in C_{j}$ are constrained to the FS sheets $C_{i,j}$.
A momentum-dependent measure of the DOS is given by the inverse Fermi velocity $1/v_{F}(\mathbf{k}) = 1/ |\nabla_{\mathbf{k}} E_{\nu} (\mathbf{k})|$, which is shown in Fig.\ \ref{fig:gap_leading_and_inverse_Fermi-velocity}(d) for the rigid bands of our model. 
Here, van Hove features can be observed on the hole-like FS near the R points. 
In order to convert the integral equation (\ref{eq:linearized_gap_equation}) to an algebraic matrix equation that can be solved numerically, the area of the discretized Fermi surface segments is determined using a Delaunay triangulation procedure \cite{kreisel_orbital_2017, durrnagel_unconventional_2022}. 

\begin{figure}[b]
\centering
\includegraphics[width=.48\textwidth]{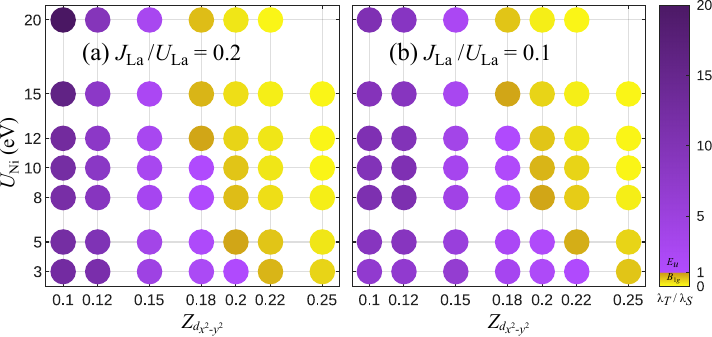}
\caption{Interaction- and renormalization-dependent leading solutions of the linearized gap equation for $U_{\text{La}}/ U_{\text{Ni}} = 0.5$ and (a) $J_{\text{La}}/U_{\text{La}} = 0.2$, (b) $J_{\text{La}}/U_{\text{La}} = 0.1$. The color scale represents the ratio $\lambda_{T} / \lambda_{S}$ of the eigenvalues of the favored triplet and singlet channel. 
The pairing symmetry is $E_{u}$ or $B_{1g}$ for a leading triplet or singlet state, respectively.}
\label{fig:phase_diagram}
\end{figure}

In Fig.\ \ref{fig:gap_leading_and_inverse_Fermi-velocity}, we present the leading gap-symmetry functions $f_{\alpha}(\mathbf{k})$ for (a) the unrenormalized and (b), (c) the renormalized scenario. Unsurprisingly, a simple \textit{d}-wave singlet solution is clearly favored for $Z_{l} = 1$, i.e., when orbital-selective renormalization is disregarded. Such a state belongs to the $B_{1g}$ irreducible representation of the point group $D_{4h}$, which describes the underlying tetragonal lattice. Here, Cooper pairing from the Ni 3$d_{x^2-y^2}$ orbital is predominant, with an almost vanishing gap on the RE electron pockets around $\Gamma$ and A. This cuprate-like SC order has been suggested previously based on spin fluctuations \cite{kitatani_nickelate_2020, sakakibara_model_2020, wu_robust_2020, kreisel_superconducting_2022, adhikary_orbital-selective_2020, lu_two-orbital_2022}.

Once we shift our model into the strongly correlated regime of $Z_{d_{x^2-y^2}} \lesssim 0.2$, solving the linearized gap equation (\ref{eq:linearized_gap_equation}) yields drastically altered results. 
The SC gap is now largest on the La dominated pockets, since the Ni $d_{x^2-y^2}$ quasiparticle DOS is strongly reduced. 
Even more importantly, two degenerate \textit{p}-wave functions become favored, including a sign change on the large FS as a function of $k_{z}$ and thus opposite signs on the pockets near $\Gamma$ and A. 
These \textit{p}-wave solutions belong to the two-dimensional representation $E_{u}$ and are likely realized in the time-reversal-symmetry-breaking (TRSB) state, for which the weak-coupling condensation energy becomes largest compared to the remaining possible combinations of the $p_{x}$ and $p_{y}$ solution \cite{sigrist_introduction_2005}. 
In the presence of weak spin-orbit coupling (see SM \cite{sup_mat}), one naturally expects an alignment between the \textit{z}-direction of the spin wave function and the crystalline \textit{c}-axis \cite{mackenzie_superconductivity_2003}. Therefore, we can analytically approximate the numerically calculated SC OP of this spin-triplet state by 
\begin{align}
    \mathbf{d}(\mathbf{k}) = \Delta_{0} \hat{\mathbf{z}} 
    \cos k_{z} 
    (\sin k_{x} \pm i \sin k_{y}),
    \label{eq:d_vector_chiral_p_wave}
\end{align}
with the magnitude of the quasiparticle gap 
\begin{align}
    \Delta_{\mathbf{k}} = |\mathbf{d}
    (\mathbf{k})| = 
    \Delta_{0} \sqrt{\cos^2 k_{z} (\sin^2 k_{x} + \sin^2 k_{y})}. 
    \label{eq:gap_magnitude}
\end{align}
Such a state is often referred to as chiral \textit{p}-wave.
Moreover, it is analogous to the so-called Anderson-Brinkman-Morel (ABM or axial) phase of $^{3}$He \cite{serene_condensed_1991, sigrist_phenomenological_1991} and has been debated extensively, e.g.,  for Sr$_{2}$RuO$_{4}$  \cite{sigrist_introduction_2005, annett_orbital_2009, mackenzie_superconductivity_2003}. 
The gap resulting from the \textbf{d} vector in (\ref{eq:d_vector_chiral_p_wave}) has point nodes on the electron pockets for $k_{x}, k_{y} = 0, \pm \pi$ and $k_{z} = \pm k_{F}$. 
In addition, line nodes are located on the large hole-like FS for $k_{x}, k_{y} = k_{F}$ and $k_{z} = \pm \pi/2$.
In Fig.\ \ref{fig:phase_diagram}, we present RPA calculations over an extensive range of input parameters to substantiate the robustness of our results.
Intriguingly, our conclusions even hold when the effect of doping is included as a rigid shift of the chemical potential such that the pocket around $\Gamma$ vanishes (see SM \cite{sup_mat}). 
One potential yet unexplored scenario involves the doping-induced transition to a distinct OP within the SC dome. 

Our proposal of chiral \textit{p}-wave superconductivity with $\mathbf{d} ~||~ \hat{\mathbf{z}}$ could explain the experimentally observed weaker suppression of superconductivity for an external magnetic field oriented within the basal plane \cite{wang_effects_2023, sun_evidence_2023, chow_pauli-limit_nodate, yan_superconductivity_2024}. For triplet superconductors, one would generally not expect Pauli suppression if $\mathbf{d}\cdot \mathbf{H} = 0$, i.e., the alignment $\mathbf{d} ~||~ \mathbf{H}$ maximizes paramagnetic limiting effects. 
Moreover, the gap in Eq.\ (\ref{eq:gap_magnitude}) has point nodes and thus could cause the quadratic temperature dependence of the London-penetration-depth \cite{smylie_evidence_2016, gordon_london_2009}, which was attributed to a superconductor with line nodes in the presence of disorder \cite{harvey_evidence_2022, chow_pairing_2023}. 
Small deviations from quadratic scaling may result from additional weight at the Fermi level, due to the horizontal line nodes on the Ni $d_{x^2-y^2}$ dominated FS with low quasiparticle DOS.
Besides, recent $\mu$SR measurements provided evidence for the coexistence of short-range magnetic order and superconductivity in IL nickelates \cite{fowlie_intrinsic_2022}, which indicates TRSB.

\textit{Summary and conclusions.}---Our calculations of magnetic fluctuations and the resulting pairing symmetries reveal potential $p_{x} + i p_{y}$ superconductivity in IL LaNiO$_{2}$, once an orbital-selective quasiparticle renormalization is introduced for the Ni $d_{x^2 - y^2}$ electrons. 
From our results, we can conclude that RE physics in LaNiO$_{2}$ is not merely important for the explanation of a negative Hall conductivity \cite{osada_nickelate_2021} but that superconductivity could actually emerge from the associated bands. 
However, one should keep in mind that the RE metal representative strongly affects the observed phenomenology of IL nickelates. 
Additionally, the proposed $p$-wave order would resolve the controversy surrounding the anisotropic high-field limiting behavior obtained from magnetotransport measurements \cite{sun_evidence_2023, chow_pauli-limit_nodate, yan_superconductivity_2024}, as well as the quadratic temperature dependence of the penetration depth \cite{smylie_evidence_2016, gordon_london_2009,harvey_evidence_2022, chow_pairing_2023}.
In this regard, our effective Hamiltonian captures the relevant physics for superconductivity in 112 lanthanum nickelates, highlighting the importance of multiorbital processes and the 3D fermiology.

To conclude, our comprehensive investigation provides strong evidence that IL LaNiO$_{2}$ could exhibit spin-triplet superconductivity, a phenomenon rarely observed. This result holds promise for a new research direction within the field of SC infinite-layer nickelates in particular and for the understanding of spin-triplet pairing in general.

\vspace{10pt}
\begin{acknowledgments}
We thank Brian M. Andersen, Matteo Dürrnagel, Ilya M. Eremin, Berit H. Goodge, Frank Lechermann, and Astrid T. Rømer for useful discussions. Financial support by the Würzburg-Dresden Cluster of Excellence ct.qmat, EXC 2147, project ID 390858490 is gratefully acknowledged. 
C. Timm acknowledges funding by the Deutsche Forschungsgemeinschaft through Collaborative Research Center SFB 1143, project A04, project ID 247310070. 
R. Thomale and A. Consiglio acknowledge support through the SFB 1170 ToCoTronics, project ID 258499086. 
A. Consiglio, D. Di Sante and R. Thomale acknowledge the Gauss Centre for Supercomputing e.V. for providing computing time on the GCS Supercomputer SuperMUC at Leibniz Supercomputing Centre. D. Di Sante received funding from the European Union Horizon 2020 research and innovation program under the Marie Sk\l odowska-Curie Grant Agreement No. 897276. A. Consiglio acknowledges support from PNRR MUR project PE0000023-NQSTI. 
\end{acknowledgments}

\bibliography{apssamp}

\end{document}


\title{Supplemental Material: Orbital-selective spin-triplet superconductivity in infinite-layer LaNiO$_{2}$} 

\author{Fabian Jakubczyk}
\email{fabian.jakubczyk@tu-dresden.de}
\affiliation{Institute of Theoretical Physics, Technische Universität Dresden, 01069 Dresden, Germany}
\affiliation{Würzburg-Dresden Cluster of Excellence ct.qmat, Germany}

\author{Armando Consiglio}
\affiliation{Würzburg-Dresden Cluster of Excellence ct.qmat, Germany}
\affiliation{Institut für Theoretische Physik und Astrophysik, Universität Würzburg, 97074 Würzburg, Germany}
\affiliation{Istituto Officina dei Materiali, Consiglio Nazionale delle Ricerche, Trieste I-34149, Italy}

\author{Domenico Di Sante}
\affiliation{Department of Physics and Astronomy,
University of Bologna, 40127 Bologna, Italy}

\author{Ronny Thomale}
\affiliation{Würzburg-Dresden Cluster of Excellence ct.qmat, Germany}
\affiliation{Institut für Theoretische Physik und Astrophysik, Universität Würzburg, 97074 Würzburg, Germany}

\author{Carsten Timm}
\affiliation{Institute of Theoretical Physics, Technische Universität Dresden, 01069 Dresden, Germany}
\affiliation{Würzburg-Dresden Cluster of Excellence ct.qmat, Germany}

\date{\today}

\maketitle

\section{DFT electronic structure}
\label{sec:DFT}
\begin{figure*}[b!]
\centering
\includegraphics[scale=0.98, trim={75 600 30 55},clip]{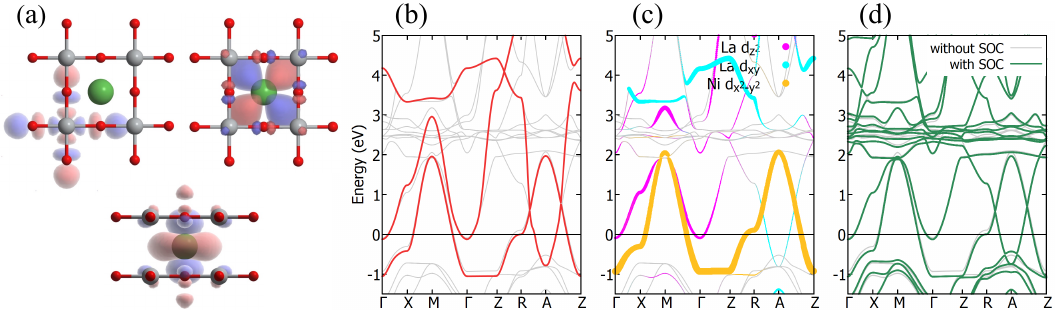}
\caption{(a) Isosurfaces of the Wannier orbitals used for the three-band model. The upper left and right panels show top views of the Ni $d_{x^2-y^2}$ and La $d_{xy}$ Wannier orbitals, respectively. The lower panel shows a side view of the La $d_{z^2}$ orbital. (b) Comparison between DFT band structure (gray lines) and bands derived from the three-band Wannier model (red lines). (c) DFT band structure projected onto the three orbitals of the model: La $d_{z^2}$, La $d_{xy}$, and Ni $d_{x^2-y^2}$. (d) Comparison between DFT band structures, with and without SOC. }
\label{fig:dft_methods}
\end{figure*}

DFT calculations have been performed using the Vienna ab-initio Simulation Package (VASP) \cite{PhysRevB.59.1758, PhysRevB.54.11169}, employing the projector augmented wave (PAW) method \cite{PhysRevB.50.17953}. Exchange and correlation effects have been handled using the generalized gradient approximation (GGA) \cite{PhysRevB.46.6671, PhysRevA.38.3098, PhysRevB.28.1809} within the Perdew-Burke-Ernzerhof (PBE) approach \cite{PhysRevLett.77.3865}. 
A plane-wave cutoff of $500\,\mathrm{eV}$ was used for the truncation of the basis set, while the relaxation of the electronic and ionic degrees of freedom was considered converged when the output difference between two steps was equal or smaller than $1\times10^{-6}\,\mathrm{eV}$ and $1\times10^{-8}\,\mathrm{eV/\AA}$, respectively. The chosen number of $\mathbf{k}$-points is 16$\times$16$\times$16. Partial occupancies have been determined via Gaussian smearing with a width of $0.1\,\mathrm{eV}$. Spin-orbit coupling (SOC) has generally not been considered, unless stated otherwise, in which case SOC has been included self-consistently. The comparison between the cases with and without SOC is shown in Fig.\ \ref{fig:dft_methods}(d).
The SOC-induced shift in energy for a given band is typically on the order of 10 to 100$\,\mathrm{meV}$. Our calculations therefore indicate that the impact of SOC is negligible also in the renormalized regime.

\begin{figure*}[t!]
\centering
\includegraphics[scale=0.225, trim={1.25cm 38cm 14.5cm 2cm},clip]{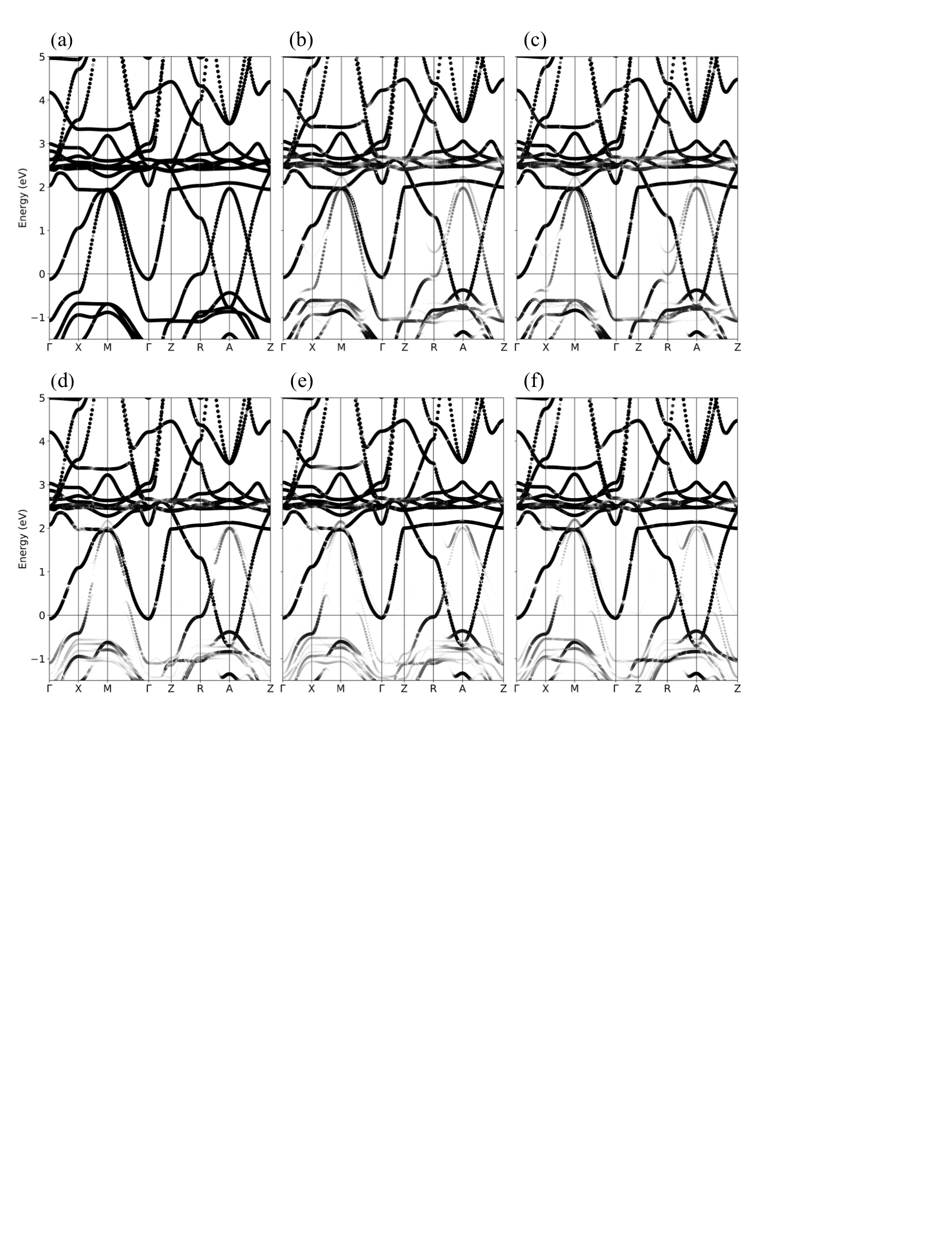}
\caption{Electronic band structures of LaNiO$_{2}$ at DFT level in several paramagnetic configurations. The weight of the dots represents the contribution of the supercell eigenstate in the primitive BZ. (a) Electronic structure of the pristine unit cell without magnetism. (b)--(f) Electronic structures of some magnetic supercells with average zero magnetization, unfolded in the primitive BZ.}
\label{fig:unfold_paramag}
\end{figure*}

In our DFT single-particle calculations, the paramagnetic phase of LaNiO$_2$ has been modeled by neglecting the spin-polarization degree of freedom. To justify this approach, and verify that the non-magnetic configuration accurately describes a paramagnetic state, we study several supercell configurations ($2\times2\times3$ expansions of the original unit cell) within the disordered local moment framework, i.e. assigning random magnetic moments to the Ni sites and ensuring that the total magnetic moment sums up to zero. This setup approximates the behavior of a real paramagnetic material. As shown in Fig.\ \ref{fig:unfold_paramag}(b) to \ref{fig:unfold_paramag}(f), the unfolded electronic band structures from several such magnetic configurations, subject to the constraints defined above, closely align with the results obtained from non-magnetic DFT calculations (Fig.  \ref{fig:unfold_paramag}(a)). This agreement supports the validity of the non-magnetic approximation for the paramagnetic state and its subsequent use in many-body analysis. For the purposes of the present work, the terms ``non-magnetic" and ``paramagnetic" can be considered as interchangeable and synonymous.

Starting from the non-magnetic system and its corresponding electronic structure, Wannier functions \cite{PhysRev.52.191, RevModPhys.34.645} have been used to provide an equivalent description of Bloch states via a maximally localized orbital basis. The Wannier-basis position matrix elements $\langle \mathbf{\mathrm{0}}n|\mathrm{\hat{r}}|\mathbf{\mathrm{R}}m \rangle$, $\mathbf{\mathrm{R}}m$ being the maximally-localized Wannier function
$m$ in unit cell $\mathbf{\mathrm{R}}$, have been used for the tight-binding (TB) description, as shown in Figs.~\ref{fig:dft_methods}(a) and \ref{fig:dft_methods}(b) \cite{MOSTOFI20142309, RevModPhys.84.1419, PhysRevB.56.12847, PhysRevB.65.035109}. 
For the three-band model, the projection functions used to build the initial guess $A(\mathbf{k})_{mn} =  \langle \psi_{m\mathbf{k}}|g_n\rangle$ for the unitary
transformation are provided by Ni $d_{x^2-y^2}$, La $d_{xy}$, and La $d_{z^2}$ orbitals. $|g_n\rangle$ represents the trial localized orbitals, and $|\psi_{m\mathbf{k}}\rangle$ are the Bloch states. 
DFT band structures have been visualized using the VASPKIT postprocessing tool, as shown in Fig.\ \ref{fig:dft_methods}(c) \cite{WANG2021108033}, while VESTA \cite{Mommako5060} has been used to visualize the Wannier-orbital isosurfaces and crystal structures in Fig.\ \ref{fig:dft_methods}(a).

\section{Wannier Hamiltonian}
\label{sec:TB_Hamiltonian}

We introduce the operator $\psi_{\sigma}^{\dagger} (\mathbf{k})= [ c_{1\sigma}^{\dagger} (\mathbf{k}), c_{2\sigma}^{\dagger}(\mathbf{k}), c_{3\sigma}^{\dagger} (\mathbf{k}) ]$ to describe the given multiband system. Here, $c_{l \sigma}^{\dagger}(\mathbf{k})$ is the fermionic creation operator, where $\sigma$ and $l$ denote the spin and orbital index, with $l=1,2,3$ referring to the La $5d_{z^2}$, La $5d_{xy}$, and Ni $3d_{x^2-y^2}$ orbital, respectively. We thus write the non-interacting Hamiltonian as
\begin{align}
    H_{0}
    = \sum_{\mathbf{k} \sigma} \psi_{\sigma}^{\dagger}(\mathbf{k}) h (\mathbf{k}) \psi_{\sigma}(\mathbf{k}). 
\label{eq:Hamiltonian_TB_momentum_general}
\end{align}

For the derivation of the matrix elements in the Hamiltonian $h(\mathbf{k})$ we employ an energy cutoff of $0.01\,\mathrm{eV}$ for the TB hopping amplitudes. The Hamiltonian is given by
\begin{widetext}
\begin{align}
h_{11} &= \epsilon_1 - \mu 
+ 2 t^{x}_{11} (\cos k_{x} + \cos k_{y})  
+ 2 t^{xx}_{11} (\cos 2 k_{x} + \cos 2 k_{y})  
+ 2 t^{z}_{11} \cos k_{z}
+ 2 t^{zz}_{11} \cos 2 k_{z} \nonumber \\
&\quad+ 4 t^{xzz}_{11} \cos 2 k_{z} (\cos k_{x} + \cos k_{y}),    \\
h_{22} &= \epsilon_2 - \mu 
    + 2 t^{x}_{22} (\cos k_{x} + \cos k_{y}) 
    + 4 t^{xy}_{22} \cos k_{x} \cos k_{y}
    + 2 t^{xx}_{22} (\cos 2 k_{x} + \cos 2 k_{y}) \nonumber \\
    &\quad+ 4 t^{xxy}_{22} (\cos 2 k_{x} \cos k_{y}+ \cos k_{x} \cos 2 k_{y}) 
    + 2 t^{z}_{22} \cos k_{z} 
    + 2 t^{zz}_{22} \cos 2 k_{z} \nonumber \\
    &\quad+ 4 t^{xz}_{22} \cos k_{z} (\cos k_{x} + \cos k_{y}) 
    + 8 t^{xyz}_{22} \cos k_{x} \cos k_{y} \cos k_{z} 
    + 4 t^{xxz}_{22} \cos k_{z} (\cos 2 k_{x} + \cos 2 k_{y}) \nonumber \\
    &\quad+ 8 t^{xxyz}_{22} \cos k_{z} (\cos 2 k_{x} \cos k_{y} 
    + \cos k_{x} \cos 2 k_{y}) 
    + 2 t^{zzz}_{22} \cos 3 k_{z} \nonumber \\
    &\quad+ 4 t^{xzz}_{22} \cos 2 k_{z} (\cos k_{x} + \cos k_{y}) 
    + 4 t^{xxzz}_{22} \cos 2 k_{z} (\cos 2 k_{x} + \cos 2 k_{y}) \nonumber \\
    &\quad+ 8 t^{xxyzz}_{22} \cos 2 k_{z} (\cos 2 k_{x} \cos k_{y} 
    + \cos k_{x} \cos 2 k_{y}),    \\ 
h_{33} &= \epsilon_3 - \mu 
    + 2 t^{x}_{33} (\cos k_{x} + \cos k_{y}) 
    + 4 t^{xy}_{33} \cos k_{x} \cos k_{y}
    + 2 t^{xx}_{33} (\cos 2 k_{x} + \cos 2 k_{y}) 
    + 2 t^{z}_{33} \cos k_{z} \nonumber \\
    &\quad+ 8 t^{xyz}_{33} \cos k_{x} \cos k_{y} \cos k_{z}, \\
h_{12} &= - 4 t^{xy}_{12} \sin k_{x} \sin k_{y} 
    - 8 t^{xyz}_{12} \sin k_{x} \sin k_{y} \cos k_{z} - 8 t^{xyzz}_{12} \sin k_{x} \sin k_{y} \cos 2 k_{z}, \\ 
h_{13} &= 8 t^{x}_{13} (\cos 3/2 k_{x} \cos 1/2 k_{y} 
    - \cos 1/2 k_{x} \cos 3/2 k_{y}) \cos 1/2 k_{z} \nonumber \\
    &\quad+ 8 t^{xz}_{13} (\cos 3/2 k_{x} \cos 1/2 k_{y} 
    - \cos 1/2 k_{x} \cos 3/2 k_{y}) \cos 3/2 k_{z},  \\   
h_{23} &= -8 t^{x}_{23} (\sin 3/2 k_{x} \sin 1/2 k_{y} 
    - \sin 1/2 k_{x} \sin 3/2 k_{y}) \cos 1/2 k_{z}  \nonumber \\
    &\quad- 8 t^{xz}_{23} (\sin 3/2 k_{x} \sin 1/2 k_{y}  
    - \sin 1/2 k_{x} \sin 3/2 k_{y})  \cos 3/2 k_{z} . 
\end{align}
\end{widetext}
We use a chemical potential of $\mu = 9.28\,\mathrm{eV}$ and $\mu = 9.12\,\mathrm{eV}$ in the undoped and doped scenario, respectively. 
The corresponding TB parameters specified in units of eV are
\begin{align}
\epsilon_{1} &= 12.3837, \; 
\epsilon_{2} = 12.3325, \; 
\epsilon_{3} = 9.6706, \;    \\ 
t^{x}_{11} &= -0.4266, \;  
t^{xx}_{11} = 0.0401, \; 
t^{z}_{11} = -1.0314, \; 
t^{zz}_{11} = 0.1150, \; 
t^{xzz}_{11} = 0.0155, \;     \\
t^{x}_{22} &= 0.3964, \;  
t^{xy}_{22} = -0.0602, \;  
t^{xx}_{22} = 0.0417, \; 
t^{z}_{22} = 0.3545, \;  
t^{xz}_{22} = -0.1680, \;  
t^{xyz}_{22} = 0.0571, \;  \nonumber \\
&\quad t^{zz}_{22} = -0.0549, \; 
t^{xxz}_{22} = -0.0242, \;  
t^{xxy}_{22} = -0.0230, \;  
t^{xxyz}_{22} = 0.0210, \;  
t^{zzz}_{22} = 0.0113, \; \nonumber \\
&\quad t^{xzz}_{22} = 0.0144, \;  
t^{xxzz}_{22} = 0.0103, \;  
t^{xxyzz}_{22} = -0.0107, \;    \\
t^{x}_{33} &= -0.3656, \;  
t^{xy}_{33} = 0.0941, \;  
t^{xx}_{33} = -0.0441, \; 
t^{z}_{33} = -0.0413, \; 
t^{xyz}_{33} = 0.0121, \;   \\
t^{xy}_{12} &= 0.0777, \;  
t^{xyz}_{12} = -0.0541, \;  
t^{xyzz}_{12} = 0.0100, \;    \\
t^{x}_{13} &= 0.0277, \;  
t^{xz}_{13} = -0.0142, \;    \\
t^{x}_{23} &= -0.0183, \;   
t^{xz}_{23} = 0.0198. \;     
\end{align}

The density of states (DOS) of the unrenormalized system can be obtained from 
\begin{align}
    \rho(\omega_{n}) = -\frac{1}{N \pi} \sum_{k} \text{Im Tr} [(i\omega_{n}\mathbb{1} - h(\mathbf{k})]^{-1}. 
\label{eq:DOS}
\end{align}

\section{renormalized susceptibilities and superconducting pairing}
\label{sec:sus}

\begin{figure*}[t]
\centering
\includegraphics[width=1\textwidth]{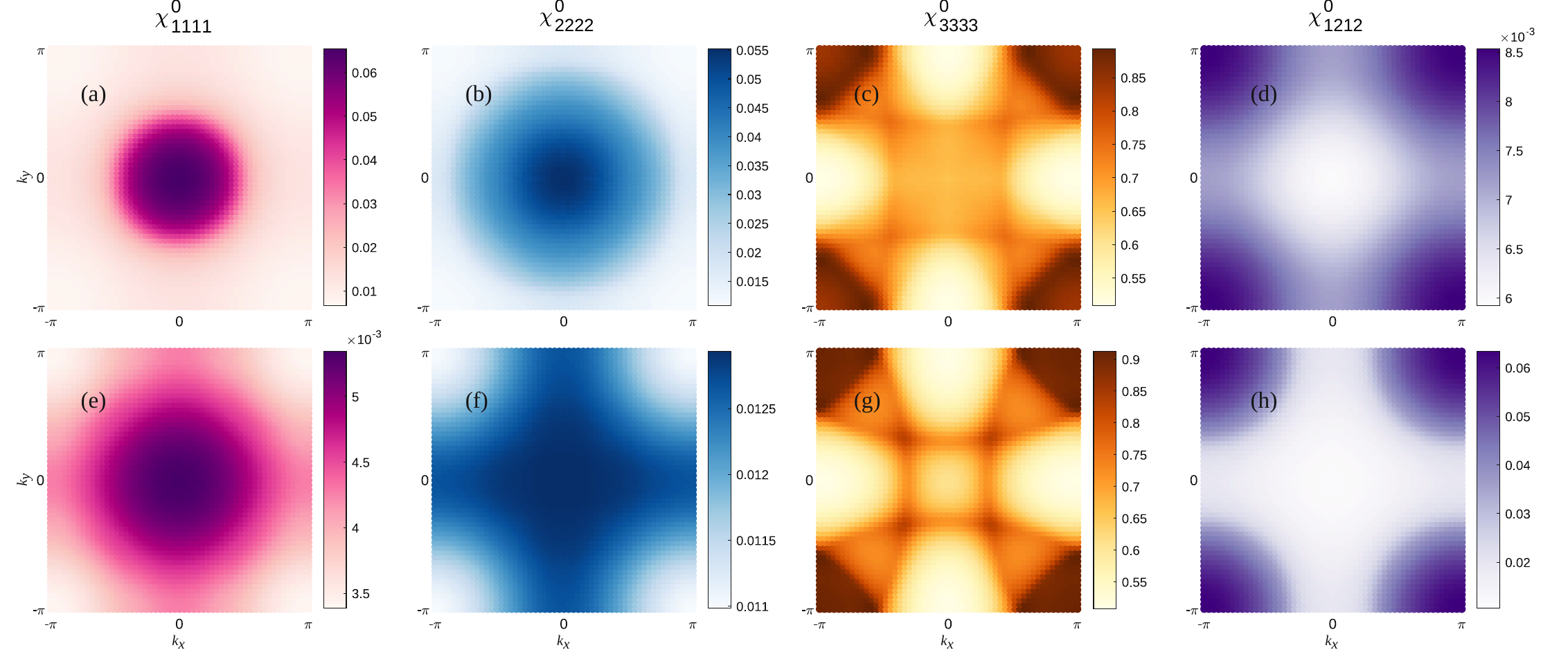}
\caption{Static ($\omega = 0$) bare susceptibilities of pristine LaNiO$_{2}$ for $Z_{d_{x^2-y^2}} = 1$ in the $k_{x}$ $k_{y}$-plane for (a) - (d) $k_{z} = 0$ and (e) - (h) $k_{z} = \pi$. 
In panels (a), (e) $\tilde{\chi}^{0}_{1111}$, (b), (f) $\tilde{\chi}^{0}_{2222}$, and (c), (g) $\tilde{\chi}^{0}_{3333}$ represent the diagonal La $d_{z^2}$, La $d_{xy}$, and Ni $d_{x^2-y^2}$ orbital contributions, respectively. Moreover, we show the interorbital susceptibility $\tilde{\chi}^{0}_{1212}$ in panels (d) and (h). 
Momentum-space integration is done on a mesh of $60 \times 60 \times 60$ \textbf{k}-points.}
\label{fig:sus_2D_cuts}
\end{figure*}

Within the orbital selective ansatz \cite{yamakawa_nematicity_2016, kreisel_orbital_2017}, quasiparticles in orbital $l$ are weighted by a factor $\sqrt{Z_{l}}$, i.e., $c_{l}^{\dagger}(\mathbf{k}) \rightarrow \sqrt{Z_{l}} c_{l}^{\dagger}(\mathbf{k})$, such that the Green's function in the orbital basis $G_{ll'}(\mathbf{k}, \omega_{n})$ becomes
\begin{align}
    \tilde{G}_{ll'}(\mathbf{k}, \omega_{n}) = 
    \sqrt{Z_{l}Z_{l'}}\sum_{\mu} \frac{a_{\mu}^{l}(\mathbf{k}) a_{\mu}^{l'*}(\mathbf{k}) }{i\omega_{n} - E_{\mu}(\mathbf{k})},
\label{eq:Greens_function_correlated}
\end{align}
where $E_{\mu} (\mathbf{k})$ is the eigenenergy of band $\mu$. 
The renormalized Green's function enters the bare 
(non-interacting) susceptibilities, which after performing Matsubara-frequency summation read as
\begin{align}
\begin{split}
    \tilde{\chi}^{0}_{l_{1}l_{2}l_{3}l_{4}} (\mathbf{q}, \omega) 
    &=-\frac{1}{N} \sum_{k, \mu \nu} \tilde{\eta}_{\mu\nu}(\mathbf{k},\mathbf{q})
    ~ 
    \frac{n_{F} (E_{\nu}(\mathbf{k}+\mathbf{q})) - n_{F} (E_{\mu}(\mathbf{k}))}{E_{\nu}(\mathbf{k}+\mathbf{q}) - E_{\mu}(\mathbf{k}) + \omega + i 0^+},  
    \label{eq:susceptibility_bare_correlated}
\end{split}
\end{align}
with the Fermi-Dirac distribution $n_{F}$. Here, we have absorbed the quasiparticle weights into the dressing factor 
\begin{align}
\begin{split}
    \tilde{\eta}_{\mu\nu}(\mathbf{k},\mathbf{q}) 
    &= \sqrt{Z_{l_{1}}Z_{l_{2}}Z_{l_{3}}Z_{l_{4}}}
    ~ 
    a_{\mu}^{l_{4}} (\mathbf{k}) a_{\mu}^{l_{2}*} (\mathbf{k}) a_{\nu}^{l_{1}} (\mathbf{k}+\mathbf{q}) a_{\nu}^{l_{3}*} (\mathbf{k}+\mathbf{q}), 
    \label{eq:dressing_factor_susceptibility_bare_correlated}
\end{split}
\end{align}
and $a_{\mu}^{l}(\mathbf{k})$ is the matrix element of the unitary transformation that connects band $\mu$ to orbital $l$. 
The homogeneous or physical bare spin susceptibility can be calculated by performing the sum  
\begin{align}
    \tilde{\chi}^{0}_{\text{phys}}(\mathbf{q}) = \frac{1}{2} \sum_{ll'} \tilde{\chi}^{0}_{lll'l'}(\mathbf{q}, 0)
    \label{eq:total_susceptibility}
\end{align}
of the static ($\omega = 0$) susceptibility \cite{graser_near-degeneracy_2009}. In Fig.\ \ref{fig:sus_2D_cuts}, we show plots of the bare susceptibilites $\tilde{\chi}^{0}_{1111}$, $\tilde{\chi}^{0}_{2222}$, $\tilde{\chi}^{0}_{3333}$, and $\tilde{\chi}^{0}_{1212}$ obtained for undoped LaNiO$_{2}$ without renormalization along 2D cuts ($k_{z} = 0, \pi$) through the Brillouin zone (BZ). 
Especially the $\tilde{\chi}^{0}_{3333}$ component shows nesting features familiar from the cuprates \cite{romer_pairing_2015}. 
However, quasiparticle renormalization strongly dampens these features, such that they become less significant in the renormalized case. 
Note that for instance the interorbital susceptibility contribution $\tilde{\chi}^{0}_{1212}$ would not be observable within the physical susceptibility given in Eq.\ (\ref{eq:total_susceptibility}). 

Interactions are included in the calculation of susceptibilities within the random phase approximation (RPA).
The spin-fluctuation and charge-fluctuation parts of the susceptibility are then given by \cite{kemper_sensitivity_2010} 
\begin{align}
    \tilde{\chi}_{l_{1}l_{2}l_{3}l_{4}}^{\text{RPA}, \,s}(\mathbf{q}, \omega) 
    &= \{\tilde{\chi}^{0}(\mathbf{q}, \omega) [\mathbb{1} - U^{s} \tilde{\chi}^{0}(\mathbf{q}, \omega)]\}_{l_{1}l_{2}l_{3}l_{4}}^{-1} ,
    \label{eq:susceptibility_RPA_spin} \\
    \tilde{\chi}_{l_{1}l_{2}l_{3}l_{4}}^{\text{RPA}, \,c}(\mathbf{q}, \omega) 
    &= \{\tilde{\chi}^{0}(\mathbf{q}, \omega) [\mathbb{1} + U^{c} \tilde{\chi}^{0}(\mathbf{q}, \omega)]\}_{l_{1}l_{2}l_{3}l_{4}}^{-1}.
    \label{eq:susceptibility_RPA_charge}
\end{align}
In orbital space, the spin and charge interaction matrices $U^{s}$ and $U^{c}$ containing the interaction parameters are given by 
\begin{align}
\begin{split}
    U^{s}_{l_{1}l_{2}l_{3}l_{4}} = \left\lbrace \begin{array}{ll}
    {U}_{\tau},  \;\;\; &l_{1} = l_{2} = l_{3} = l_{4}, \\ 
    {U}_{\tau}', \;\;\; &l_{1} = l_{3} \neq l_{2} = l_{4}, \\
    {J}_{\tau},  \;\;\; &l_{1} = l_{2} \neq l_{3} = l_{4}, \\
    {J}_{\tau}', \;\;\; \;\;\; \;\;\; \;\;\; \;\;\; \; &l_{1} = l_{4} \neq l_{2} = l_{3}, 
    \end{array}
    \right.
\end{split} \\
\begin{split}
    U^{c}_{l_{1}l_{2}l_{3}l_{4}} = \left\lbrace \begin{array}{ll}
    {U}_{\tau},  \;\;\; &l_{1} = l_{2} = l_{3} = l_{4}, \\ 
    -{U}_{\tau}' + 2{J}_{\tau}, \;\;\; &l_{1} = l_{3} \neq l_{2} = l_{4}, \\
    2{U}_{\tau}' - {J}_{\tau},  \;\;\; &l_{1} = l_{2} \neq l_{3} = l_{4}, \\
    {J}_{\tau}', \;\;\; &l_{1} = l_{4} \neq l_{2} = l_{3}, 
    \end{array}
    \right.
\end{split}
\end{align}
respectively, where $\tau \in (\text{La, Ni})$ is the sublattice index. 

In order to determine the superconducting (SC) pairing symmetries, we solve the linearized gap equation 
\begin{align}
    \lambda_{\alpha} f_{\alpha}(\mathbf{k}) = 
    \sum_{j} \oint_{C_{j}} \frac{\text{d}^2 k_{\|}'}{(2\pi)^2 v_{F}(\mathbf{k}')} \tilde{\Gamma}_{ij}(\mathbf{k},\mathbf{k'}) f_{\alpha}(\mathbf{k}').
\label{eq:linearized_gap_equation}
\end{align}
The largest eigenvalue $\lambda_{\alpha}$ of equation (\ref{eq:linearized_gap_equation}) results in the highest transition temperature \cite{graser_near-degeneracy_2009}, while $\lambda_{\alpha}$ in general determines the pairing strength for the different pairing channels $\alpha$ \cite{scalapino_d_1986}. The corresponding gap symmetries are identified by the eigenfunction $f_{\alpha}(\mathbf{k})$. 
The inverse of the Fermi velocity $v_{F}(\mathbf{k}) = |\nabla_{\mathbf{k}} E_{\nu} (\mathbf{k})|$ gives a measure of the DOS and momenta $\mathbf{k} \in C_{i}$, $\mathbf{k}' \in C_{j}$ are constrained to the FS sheets $C_{i,j}$. 
In Eq.\ (\ref{eq:linearized_gap_equation}), $\tilde{\Gamma}_{ij}(\mathbf{k},\mathbf{k'})$ represents the renormalized form of the effective multiorbital pair scattering vertex and is given by
\begin{widetext}
\centering
\begin{align}
    \tilde{\Gamma}_{ij}(\mathbf{k},\mathbf{k'}) 
    &= \text{Re}  \sum_{l_{1}l_{2}l_{3}l_{4}} 
    \sqrt{Z_{l_{1}}}\sqrt{Z_{l_{4}}}
    a_{\nu_{i}}^{l_{1},*}(\mathbf{k)} 
    a_{\nu_{i}}^{l_{4},*}(\mathbf{-k)} 
    ~ \tilde{\Gamma}_{l_{1}l_{2}l_{3}l_{4}}^{S/T} (\mathbf{k, k'}, \omega=0)
    ~ 
    \sqrt{Z_{l_{2}}}\sqrt{Z_{l_{3}}}
    a_{\nu_{j}}^{l_{2}}(\mathbf{k')} 
    a_{\nu_{j}}^{l_{3}}(\mathbf{-k'}), \\
    \tilde{\Gamma}_{l_{1}l_{2}l_{3}l_{4}}^{S}(\mathbf{k},\mathbf{k'},\omega) 
    &= \left[ \frac{3}{2} U^{s} \tilde{\chi}_{l_{1}l_{2}l_{3}l_{4}}^{\text{RPA}, \,s}(\mathbf{k-k'})U^{s} + \frac{1}{2}U^{s} -\frac{1}{2}U^{c} \tilde{\chi}_{l_{1}l_{2}l_{3}l_{4}}^{\text{RPA}, \,c}(\mathbf{k-k'})U^{c} + \frac{1}{2}U^{c} \right] _{l_{1}l_{2}l_{3}l_{4}},\\
    \tilde{\Gamma}_{l_{1}l_{2}l_{3}l_{4}}^{T}(\mathbf{k},\mathbf{k'},\omega) 
    &= \left[ -\frac{1}{2} U^{s} \tilde{\chi}_{l_{1}l_{2}l_{3}l_{4}}^{\text{RPA}, \,s}(\mathbf{k-k'})U^{s} + \frac{1}{2}U^{s} -\frac{1}{2}U^{c} \tilde{\chi}_{l_{1}l_{2}l_{3}l_{4}}^{\text{RPA}, \,c}(\mathbf{k-k'})U^{c} + \frac{1}{2}U^{c} \right] _{l_{1}l_{2}l_{3}l_{4}} .
\end{align}
\end{widetext}
Here, $\tilde{\Gamma}_{l_{1}l_{2}l_{3}l_{4}}$ determines the orbital vertex functions in the singlet (\textit{S}) and triplet (\textit{T}) channel. 
Physically, $\tilde{\Gamma}_{l_{1}l_{2}l_{3}l_{4}}$ reflects particle-particle scattering of electrons from orbitals $l_{1}$, $l_{4}$ into orbitals $l_{2}$, $l_{3}$. The quantities $U^{s}$, $U^{c}$, $\chi^{\text{RPA}}$ are matrices in orbital space and $\chi^{\text{RPA}, s/c}$ describes spin or charge fluctuations, respectively.

\section{Stability of RPA analysis with respect to TB parameters}
\label{sec:stability_RPA_TB}

\begin{figure}[b!]
\centering
\includegraphics[width=.95\textwidth]{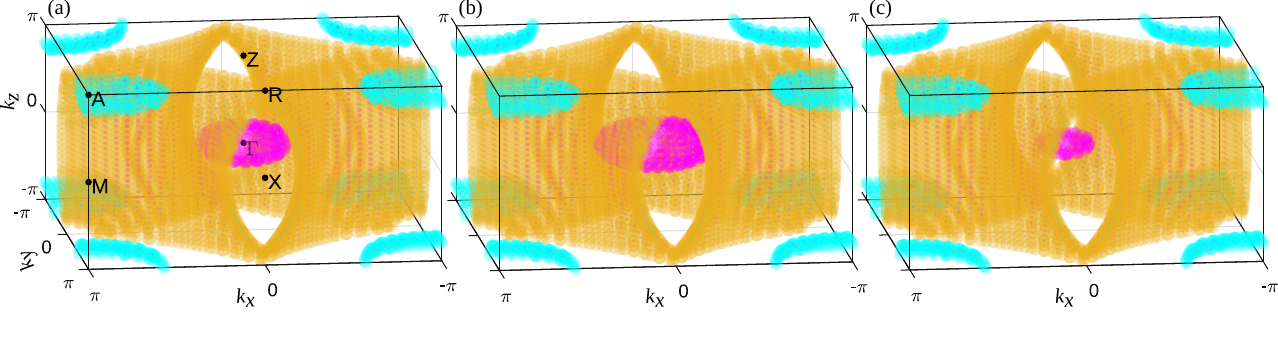}
\caption{3D Fermi surfaces, with magenta, cyan, and yellow color referring to La $d_{z^2}$, La $d_{xy}$, and Ni $d_{x^2-y^2}$ orbital contributions, respectively. Shown are (a) the FS from the unchanged TB model, (b) the results from increasing $t^z_{11}$ by $5\%$, and (c) from reducing $t^z_{11}$ by $5\%$.}
\label{fig:FS_3D_orbitally_mu_9.28_100_105_095_tz11}
\end{figure}

\begin{table}[b!]
\centering
\setlength{\extrarowheight}{3pt} 
\begin{tabular*}{.95\linewidth}{@{\extracolsep{\fill}} 
c | c c c c c c c c c c c}
$U_{\text{Ni}}$ & unchanged & $0.95 \; t^x_{11}$ & $1.05 \; t^x_{11}$ & $0.95 \; t^z_{11}$ & $1.05 \; t^z_{11}$ & $0.95 \; t^x_{22}$ & $1.05 \; t^x_{22}$ & $0.95 \; t^z_{22}$ & $1.05 \; t^z_{22}$ & $0.95 \; t^x_{33}$ & $1.05 \; t^x_{33}$ \\
\hline \hline
5 & 3.67 & 3.16 & 3.73 & 2.95 & 3.69 & 3.47 & 3.30 & 3.90 & 3.44 & 3.71 & 3.56 \\ 
\arrayrulecolor{lightgray}
\hline 
\arrayrulecolor{black}
20 & 2.53 & 1.85 & 3.96 & 1.75 & 4.81 & 2.48 & 2.41 & 2.55 & 2.47 & 2.48 & 2.50 \\ 
\end{tabular*}
\caption{Ratio $\lambda_{T}/\lambda_{S}$ of the eigenvalues of the favored triplet and singlet channel for $U_{\text{La}}/ U_{\text{Ni}} = 0.5$, $J_{\text{La}}/U_{\text{La}} = 0.2$, and $Z_{d_{x^2-y^2}} = 0.15$. 
The pairing symmetry remains $E_{u}$ throughout.}
\label{table:lambda_ratio_change_hoppings}
\end{table}

A recent debate regarding the pairing symmetry in the La$_{3}$Ni$_{2}$O$_{7}$ nickelate system \cite{liu_s_2023, heier_competing_2024} stresses the importance of potential sensitivity of the RPA method to details of the input band structure and interaction parameters. In order to ensure the robustness of our results not just with respect to quasiparticle renormalization and interaction strength, which we discuss in the main text, we also investigate the impact of small modification to the TB model. 
When doing so, our aim is to keep the overall FS topology close to the one of the unchanged TB model. This results in a limitation of the variation of the hopping amplitudes. If, for instance, we vary the parameter $t^z_{11}$ by $\pm 5 \%$, the pocket around $\Gamma$ notably changes in size, as can be seen in Fig.\ \ref{fig:FS_3D_orbitally_mu_9.28_100_105_095_tz11}. 
For larger changes, it would eventually vanish or touch the Ni $3d_{x^2-y^2}$ dominated surface, which represents a major deviation from the original fermiology. We take this as a point of reference and calculate the paring symmetries for $\pm 5 \%$ changes to the largest hopping amplitudes, i.e., $t^x_{11}$, $t^z_{11}$, $t^x_{22}$, $t^z_{22}$, and $t^x_{33}$.

As a result, we find that in all cases the leading renormalized solution still remains a degenerate $(p_{x}, p_{y})$-wave state. In Table \ref{table:lambda_ratio_change_hoppings}, we list the corresponding ratio $\lambda_{T}/\lambda_{S}$ of leading triplet over singlet eigenvalue for $U_{\text{La}}/ U_{\text{Ni}} = 0.5$, $J_{\text{La}}/U_{\text{La}} = 0.2$, $U_{\text{Ni}} = 5\,\mathrm{eV}$ and $20\,\mathrm{eV}$, as well as $Z_{d_{x^2-y^2}} = 0.15$. 
Another modification that leaves the general FS topology intact would be given by alteration of the interorbital, i.e., hybridization terms in the Hamiltonian. Here, we checked the extreme case of essentially neglecting hybridization by setting all $t_{12}$, $t_{13}$, or $t_{23}$ amplitudes to zero, both individually or together. Exemplary calculations with $J_{\text{La}}/U_{\text{La}} = 0.2$ and $U_{\text{Ni}} = 5\,\mathrm{eV}$ show negligible changes of $\lambda_{T}/\lambda_{S}$, while confirming the degenerate $p$-wave solution.
Hence, we conclude that our DFT band structure indeed permits a reliable RPA analysis and physically reasonable modifications to the TB model do not significantly alter the leading terms of the pairing symmetries.

\section{Doped {L\lowercase{a}N\lowercase{i}O}$_{2}$}
\label{sec:doped_LaNiO2}

\begin{figure}[t!]
\centering
\includegraphics[width=.95\textwidth]{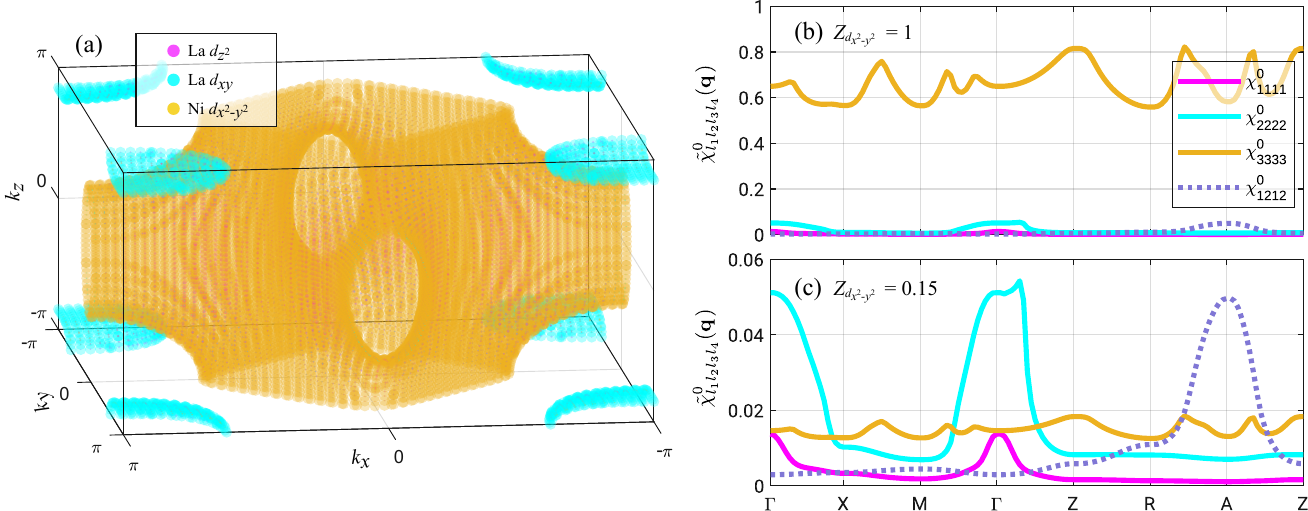}
\caption{(a) 3D Fermi surface including the orbital contributions relevant for the low-energy description of doped LaNiO$_{2}$. Magenta, cyan, and yellow refer to the La $d_{z^2}$, La $d_{xy}$, and Ni $d_{x^2-y^2}$ orbitals, respectively. (b), (c) Static bare susceptibilities $\tilde{\chi}^{0}_{l_{1}l_{2}l_{3}l_{4}} (\mathbf{q}, \omega = 0) $ for (b) $Z_{d_{x^2-y^2}} = 1$ and (c) $Z_{d_{x^2-y^2}} = 0.15$ for doped LaNiO$_{2}$. The interorbital susceptibility $\tilde{\chi}^{0}_{1212}$ is shown in purple. The mesh for momentum space integration includes $80 \times 80 \times 80$ \textbf{k}-points.}
\label{fig:3D_FS_and_sus_doped}
\end{figure}

\begin{figure*}[b!]
\centering
\includegraphics[width=1\textwidth]{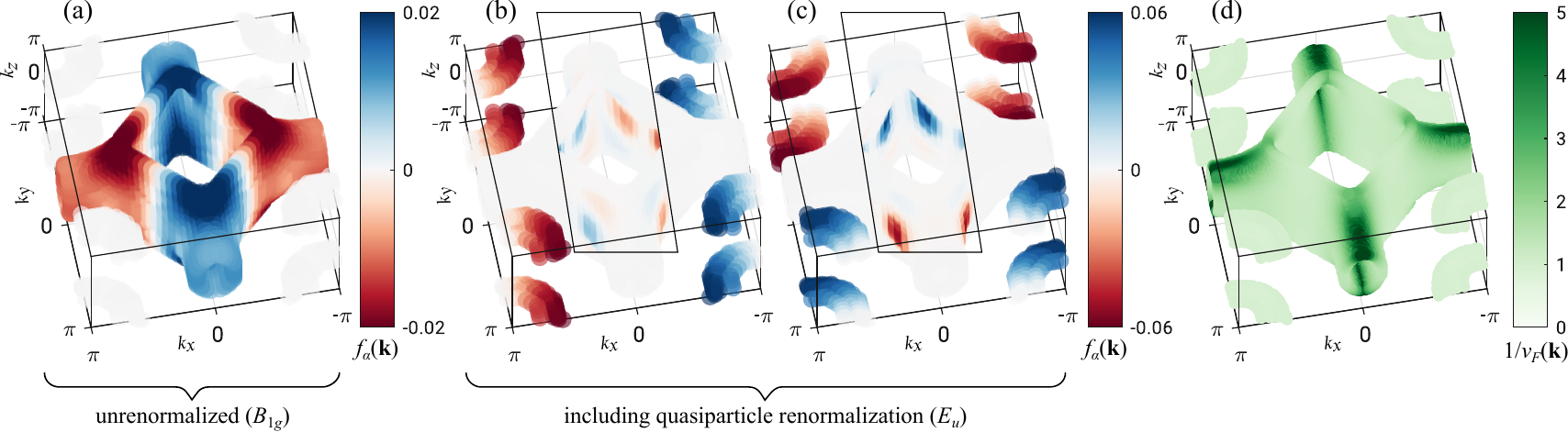}
\caption{Leading pairing structure of LaNiO$_{2}$ (a) without and (b), (c) with inclusion of quasiparticle renormalization.
The gap symmetry changes from (a) an even-parity $d$-wave to (b), (c) a degenerate odd-parity $p$-wave solution. 
A color scale highlighting smaller amplitudes with maximum value of $\pm 1 \times 10^{-4}$ was chosen for the insets in panels (b), (c).
(d) Inverse Fermi-velocity $1/v_{F}(\mathbf{k}) = 1/|\nabla_{\mathbf{k}} E_{\nu} (\mathbf{k})|$ giving a measure of the DOS. 
A total of 6800 points is taken for the 3D gap-structure calculations and we use $60 \times 60 \times 60$ \textbf{k}-points for momentum space integration.}
\label{fig:gap_leading_and_inverse_Fermi-velocity_doped}
\end{figure*}

Another natural question is whether the observed odd-parity \textit{p}-wave state is also realized in doped LaNiO$_{2}$. To investigate this, we capture the essential impact of doping by a rigid shift of the chemical potential, such that the $\Gamma$-centered electron pocket is lifted above the Fermi level. We show the corresponding 3D FS in Fig.\ \ref{fig:3D_FS_and_sus_doped}(a). 
In the next step, we analyze the bare susceptibilities shown in Figs.\ \ref{fig:3D_FS_and_sus_doped}(b) and \ref{fig:3D_FS_and_sus_doped}(c). 
On the one hand, the diagonal susceptibility $\tilde{\chi}^{0}_{3333}$ looses its main peaks around the A and M points.  
On the other hand, once we introduce correlations by setting $Z_{d_{x^2-y^2}} = 0.15$, the central features are again given by the intraorbital rare-earth contributions $\tilde{\chi}^{0}_{1111}$ and $\tilde{\chi}^{0}_{2222}$ around $\Gamma$, as well as the interorbital component $\tilde{\chi}^{0}_{1212}$ around A. In contrast to undoped LaNiO$_{2}$, the intraorbital La $d_{xy}$ susceptibility $\tilde{\chi}^{0}_{2222}$ now far surpasses $\tilde{\chi}^{0}_{1111}$ from the La $d_{z^2}$ orbital. 
This is not surprising since La $d_{z^2}$ electrons used to contribute mainly to the pocket around $\Gamma$, which has vanished due to doping. 

Finally, we also perform calculations of the pairing symmetry in doped LaNiO$_{2}$, which can be seen in Fig.\ \ref{fig:gap_leading_and_inverse_Fermi-velocity_doped}. 
Most interestingly, we observe that the essential conclusions made for the SC state of the undoped compound are still valid. The gap now clearly dominates on the remaining rare-earth pockets around A. 
In fact, the transition to a favored \textit{p}-wave happens for even bigger quasiparticle weights of $Z_{d_{x^2-y^2}} \lesssim 0.3$, which might be facilitated due to a reduced DOS at the van Hove features around R.  
For $Z_{d_{x^2-y^2}} = 0.15$, $U_{\text{La}}/ U_{\text{Ni}} = 0.5$, $J_{\text{La}}/U_{\text{La}} = 0.2$ and $U_{\text{Ni}} = 5\,\mathrm{eV}$, the eigenvalue $\lambda_{T}$ of the degenerate leading triplet solutions, i.e., their pairing strength, reaches $\lambda_{T} \approx 8.99\lambda_{S}$, with $\lambda_{S}$ belonging to the leading singlet state. Intriguingly, this ratio is even greater than in pristine LaNiO$_{2}$, where $\lambda_{T} \approx 3.67\lambda_{S}$ for the same set of parameters.
However, note that doping reduces the effective-mass enhancement and thus could have a compensating effect.

\bibliography{SM_apssamp}